# Self-Induced Decay of Intense Laser Pulse into a Pair of Surface Plasmons

*Ivan Oladyshkin*

*Institute of Applied Physics of the Russian Academy of Sciences, 603950, Nizhny Novgorod, Russia*

*oladyshkin@ipfran.ru*

We show theoretically that an intense femtosecond optical pulse incident normally on a metal surface tends to decay into a pair of counter-propagating surface plasmon-polaritons (SPPs). The interference field heats the medium periodically, which causes a periodic permittivity perturbation and resonantly amplifies the magnitudes of SPPs. The instability growth time is only 10–50 fs for typical metals at damaging laser fluences. This mechanism is promising for the interpretation of laser-induced periodic surface structures formation in a single-pulse pumping regime.

The phenomenon of laser-induced periodic surface structures (LIPSS) formation has been studied for more than 50 years since the first observation by M. Birnbaum in Ge and GaAs samples in 1965 [1]. Further it was found that the irradiation of metals and dielectrics by normally incident short laser pulses of near-threshold fluence also leads to appearing of periodic ripples on the surface (e.g., see recent review papers [2, 3]). In addition to the fundamental interest, this effect attracts attention as relatively simple and fast method of submicron material structuring. An orientation and a period of ripples are determined both by the material properties and the laser irradiation parameters; the detailed analysis of LIPSS characteristics can be found in Refs. [2–4]. Below we focus only on the case of metals and highly doped semiconductors, where the condition of surface plasmon-polariton existence is fulfilled, i.e., the real part of permittivity $\varepsilon'$ is less than –1.

The key role of SPP in the formation of LIPSS in metals has been proven in many experimental and theoretical papers [2-7]. The period of ripples is usually equal to the SPP period or twice smaller. It was also shown that the pre-structuring of the surface for better coupling between the incident wave and SPP increases further growth of ripples [8, 9]. However, during the multi-pulse machining and the growth of LIPSS depth their period gradually decreases which is probably a consequence of the efficient permittivity change in the damaged layer.

General electromagnetic theory of LIPSS formation assisted by SPP excitation was developed by Sipe and co-authors in early 80-s in the series of pioneer papers [5–7]. In the framework of this model the SPP (along with other diffraction waves) are generated on random surface irregularities and simultaneously interfere with the incident laser pulse. The resulting interference pattern has the period equal to the SPP wavelength which explains the period of LIPSS observed in metals experimentally. The mentioned series of papers by Sipe became the base for the developing of various more complicated and detailed models of LIPSS formation [10-13].

Despite the undoubted progress made in this field, the very initial stage of LIPSS formation along with the single-pulse experiments remain not fully understood from the theoretical point of view, which was discussed in details in the review paper [2]. In a nutshell, we cannot expect enough efficient excitation of SPP on a random surface. Metal and semiconductor samples used in real experiments are rather smooth, and so the spectral measure of their surface roughness $b(\mathbf{k})$ introduced by Sipe [5] is very small in the resonant interval of wavenumbers $\mathbf{k}$. For comparison, the influence of the surface pre-structuring on LIPSS formation was proved experimentally in Refs. [8] and [9] for the depths of *resonant* gratings of 10 nm and 66 nm, respectively. Another close example is the optical-to-terahertz conversion in metals: it was shown theoretically and experimentally that the significant influence of SPPs on the terahertz output takes places when the depth of a resonant grating is more than 20–30 nm [14, 15]. Obviously, if we consider not a regular, but a random surface shape, it should have much larger amplitude of irregularities to reach comparable SPP excitation efficiency. However, such rough samples are not used in the experiments on LIPSS formation. Also note that in recent theoretical papers (see [11-13] and many others) the efficiency of SPP generation is usually chosen as a constant free parameter.

In this Letter we show that the surface roughness is not necessary the main source of surface plasmons, while it just creates the initial conditions for SPP nonlinear growth during the laser pulse action (the alternative sources of seed SPPs are thermal fluctuations of the electron density and the electromagnetic field). To describe it theoretically, we should take into account that the interference pattern of SPPs and the incident wave not only influences on the further material melting, but also causes periodic heating of electrons which changes local permittivity strongly. Below we demonstrate that the discussed feedback mechanism leads to the instability of a flat wave reflection with respect to decay into a pair of counter-propagating SPPs. The instability growth time (estimated for the set of typical parameters of gold) may be as short as 10–50 fs for the incident fluences of about 1 J/cm$^2$ used in real single-shot experiments. It means that during the femtosecond laser pulse action, the SPP magnitude has enough time to growth several orders higher than it would be expected from the theory of linear transformation on random irregularities. To take this into account, we describe the evolution of SPP magnitude explicitly, without introducing any phenomenological transformation coefficients.

**SPPs excitation on the permittivity perturbation.** Let us start by considering the influence of some permittivity perturbation $\delta\varepsilon$ on the reflection process. Since we assume that the perturbations are relatively small, the problem of the laser pulse diffraction can be linearized, so each Fourier component of the incident pulse and of $\delta\varepsilon$ can be taken into account independently. The geometry of the problem is shown in Fig. 1: a flat conductor with the permittivity $\varepsilon$ ($Re\ \varepsilon < -1$) occupies the region $x < 0$ and the optical wave with the wavenumber $k_0$ and frequency $\omega$ polarized along the $z$-axis incidents normally at its surface. In the absence of any medium perturbations the electric field above and below the surface is described as the sum of three waves:

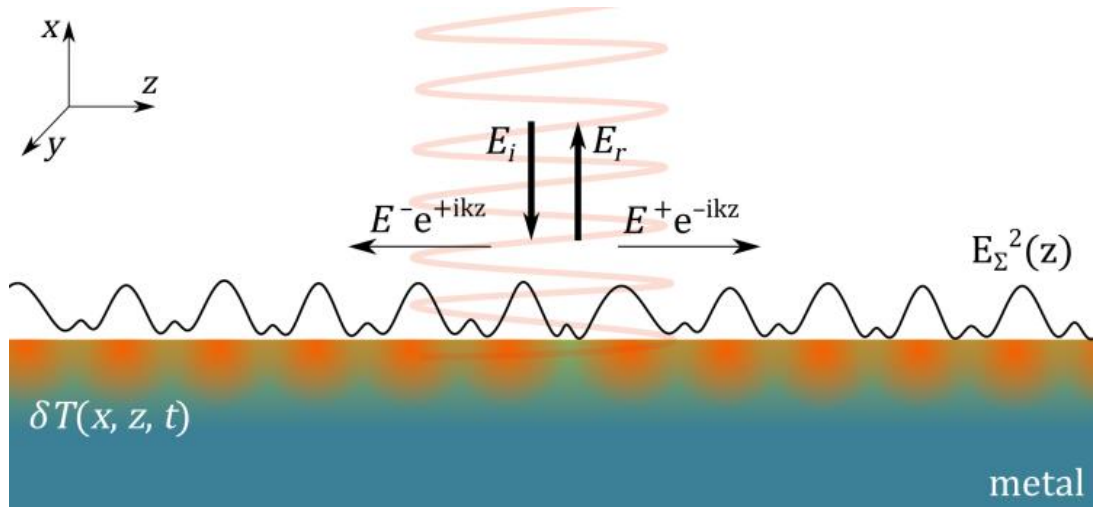

Fig 1. Incident wave decay into a pair of SPPs. The resulting optical field intensity at the surface $\mathbf{E}_\Sigma^2(z)$ is shown by a black curve. Periodic temperature perturbation inside the metal is shown by a color gradient.

$$x > 0: \quad \mathbf{E}_i(t, x, z) = \mathbf{z_0} E_i \exp(i\omega t + ik_0 x), \tag{1}$$

$$x > 0: \quad \mathbf{E}_r(t, x, z) = \mathbf{z_0} E_r \exp(i\omega t - ik_0 x), \tag{2}$$

$$x < 0: \quad \mathbf{E}_t(t, x, z) = \mathbf{z_0} E_t \exp(i\omega t + \alpha x), \tag{3}$$

where $\mathbf{z}_0$ is the unit vector, $\alpha = \sqrt{-\varepsilon k_0^2}$, $\mathbf{E}_i$, $\mathbf{E}_r$ and $\mathbf{E}_t$ are the electric fields of the incident, reflected and transmitted waves, respectively, with the magnitudes determined by the Fresnel relations:

$$E_t = \frac{2E_i}{1-\sqrt{\varepsilon}}, \quad E_r = E_i \frac{1+\sqrt{\varepsilon}}{1-\sqrt{\varepsilon}}. \tag{4}$$

Now let us move to a permittivity perturbation of the harmonic form with an arbitrary wavenumber $k_\varepsilon$, so $\varepsilon = \varepsilon_0 + \delta\varepsilon(x, z, t)$ and

$$\delta\varepsilon(x, z, t) = \delta\tilde{\varepsilon}(x, t) \cdot \cos k_\varepsilon z, \tag{5}$$

where $\delta\tilde{\varepsilon}(x, t)$ assumed to be slowly varying function of time ($|\partial\delta\tilde{\varepsilon}/\partial t| \ll \omega\delta\tilde{\varepsilon}$) and the characteristic spatial scale of $\delta\tilde{\varepsilon}(x)$ is expected to be comparable with the optical skin layer. Note that an arbitrary phase of the perturbation $\sim \cos(k_\varepsilon z + \varphi)$ would not change further analysis. Evolution of $\delta\tilde{\varepsilon}$ due to the heating of electrons will be considered after solving the diffraction problem.

Substituting an inhomogeneous permittivity to Maxwell equations, we obtain the following equation inside the conductor for the total magnetic field $\mathbf{H}$ and electric field $\mathbf{E}$:

$$\Delta \mathbf{H} - \frac{\varepsilon}{c^2} \frac{\partial^2}{\partial t^2} \mathbf{H} = \frac{1}{c} \frac{\partial}{\partial t} [\nabla\varepsilon, \mathbf{E}], \tag{6}$$

where $\Delta$ and $\nabla$ are the Laplace and nabla differential operators respectively, and the square brackets denote the cross product. Using perturbation theory, we can separate diffraction fields in the explicit form representing the solution of Eq. (6) as a sum of unperturbed fields inside the medium $\mathbf{E}_t$ and $\mathbf{H}_t$ and the first order fields $\mathbf{E}_1$ and $\mathbf{H}_1$: $\mathbf{E} = \mathbf{E}_t + \mathbf{E}_1$ and $\mathbf{H} = \mathbf{H}_t + \mathbf{H}_1$. Neglecting second order terms we obtain:

$$\Delta \mathbf{H}_1 - \frac{\varepsilon_0}{c^2} \frac{\partial^2 \mathbf{H}_1}{\partial t^2} = \frac{1}{c} \frac{\partial}{\partial t} [\nabla\delta\varepsilon, \mathbf{E}_t] + \frac{\delta\varepsilon}{c^2} \frac{\partial^2 \mathbf{H}_t}{\partial t^2}, \tag{7}$$

which is a common wave equation with a source. In the chosen geometry both terms in the right side of Eq. (7) are directed along the $y$-axis, so the magnetic field has only a $y$-component: $\mathbf{H}_1 = H_1 \mathbf{y}_0$. Since Eq. (7) is linear with respect to the electromagnetic fields and the permittivity varies relatively slowly, the frequency transformation effects are negligible and the diffraction fields can be represented as the harmonic waves $\sim e^{i\omega t}$, which leads to

$$\Delta H_1 + \varepsilon_0 k_0^2 H_1 = -ik_0 E_t e^{\alpha x} \frac{\partial \delta\varepsilon}{\partial x} - k_0^2 \delta\varepsilon H_t e^{\alpha x}. \tag{8}$$

The right side of Eq. (8) contains only two spatial harmonics $e^{\pm ik_\varepsilon z}$. Taking this into account and writing the magnetic field $H_1(x, z, t)$ as

$$H_1(x, z, t) = H_1^+(x) e^{i\omega t - ik_\varepsilon z} + H_1^-(x) e^{i\omega t + ik_\varepsilon z}, \tag{9}$$

we come to similar equations for both spatial harmonics $H_1^\pm$ (which meets the requirements of symmetry $z \to -z$):

$$\frac{\partial^2 H_1^\pm}{\partial x^2} - k_\varepsilon^2 H_1^\pm + \varepsilon_0 k_0^2 H_1^\pm = -\frac{ik_0 E_t e^{\alpha x}}{2} \frac{\partial \delta\tilde{\varepsilon}(x)}{\partial x} - \frac{k_0^2 \delta\tilde{\varepsilon}(x)}{2} H_t e^{\alpha x}. \tag{10}$$

Eq. (10) can be solved analytically for some model profiles $\delta\tilde{\varepsilon}(x)$; from further analysis it follows that the specific form of this function influences the effect we focus on insignificantly. Assuming that $\delta\tilde{\varepsilon}$ should decrease with depth we set it as $\delta\tilde{\varepsilon}(x) = \delta\tilde{\varepsilon}\, e^{gx}$ and obtain the following solution for the magnetic field:

$$x < 0:\ H_1^{\pm} = A e^{\alpha_2 x} + \Psi(x), \tag{11}$$

where $A$ is an arbitrary constant, $\alpha_2 = \sqrt{k_\varepsilon^2 - \varepsilon_0 k_0^2}$ and

$$\Psi(x) = -\frac{ik_0 g - k_0^2 \sqrt{\varepsilon_0}}{(g+\alpha)^2 - k_\varepsilon^2 + \varepsilon_0 k_0^2} \frac{E_i \delta\tilde{\varepsilon}}{(1-\sqrt{\varepsilon_0})} e^{(g+\alpha)x}. \tag{12}$$

The magnetic field in free space $x > 0$ also consists of two spatial harmonics $H_1^{\pm}(x) e^{i\omega t \mp i k_\varepsilon z}$, which, according to Helmholtz equation, gives

$$x > 0:\ H_1^{\pm}(x) = C e^{-\alpha_1 x}, \tag{13}$$

where $\alpha_1 = \sqrt{k_\varepsilon^2 - k_0^2}$ and $C$ is another arbitrary constant. Note that an imaginary value of $\alpha_1$ ($k_\varepsilon < k_0$) does not contradict our consideration and corresponds to the case of propagating scattered waves in the upper half-space, but the chosen notation is more natural for the analysis of localized plasmon modes.

To find both constants $A$ and $C$, the boundary conditions should be taken into account. First, the amplitudes of tangential electric fields of both harmonics $E_z^{\pm}$ can be found from Maxwell equations:

$$x < 0:\ ik_0 \varepsilon E_z^{\pm} = \frac{\partial H_1^{\pm}(x)}{\partial x}, \tag{14}$$

$$x > 0:\ ik_0 E_z^{\pm} = \frac{\partial H_1^{\pm}(x)}{\partial x}. \tag{15}$$

Using the conditions of the magnetic and electric field continuity at the boundary $x = 0$, we find:

$$A = \frac{\varepsilon_0 \alpha_1 + g + \alpha}{\alpha_2 + \varepsilon_0 \alpha_1} \frac{ik_0 g - k_0^2 \sqrt{\varepsilon_0}}{(g+\alpha)^2 - k_\varepsilon^2 + \varepsilon_0 k_0^2} \frac{E_i \delta\tilde{\varepsilon}}{(1-\sqrt{\varepsilon_0})}, \tag{16}$$

$$C = \frac{g + \alpha - \alpha_2}{\alpha_2 + \varepsilon_0 \alpha_1} \frac{ik_0 g - k_0^2 \sqrt{\varepsilon_0}}{(g+\alpha)^2 - k_\varepsilon^2 + \varepsilon_0 k_0^2} \frac{E_i \delta\tilde{\varepsilon}}{(1-\sqrt{\varepsilon_0})}. \tag{17}$$

Both of the obtained expressions (16)–(17) have the same resonant denominator $D = \alpha_2 + \varepsilon_0 \alpha_1$, which reaches the minimal value when the wavenumber of permittivity perturbation $k_\varepsilon$ coincides with the real part of SPP wavenumber:

$$k_\varepsilon = k_0 \sqrt{\frac{\varepsilon_0'}{1+\varepsilon_0'}}, \tag{18}$$

where $\varepsilon_0 = \varepsilon_0' + i\varepsilon_0''$. Note that near the resonant point $H_1^{\pm}(x=0) \cong A \cong C \gg \Psi(x=0)$.

Eqns. (16)–(17) give only the stationary amplitude of SPP generated by a continuous monochromatic incident wave and limited by the absorption effects ($Im[D] \propto i\varepsilon_0''$). To describe the dynamics of SPP excitation by a femtosecond laser pulse, one needs to consider the spectral properties of Eqns. (16)–(17), for which some specific model of the medium permittivity $\varepsilon(\omega)$ should be introduced. Since the aim of this Letter is to demonstrate fundamental feasibility of spontaneous decay, we use the simplest Drude model:

$$\varepsilon(\omega) = 1 - \frac{\omega_p^2}{\omega(\omega - i\nu)} \cong -\frac{\omega_p^2}{\omega^2} - i\frac{\nu\omega_p^2}{\omega^3}, \tag{19}$$

where $\omega_p$ is the plasma frequency, $\nu$ is the scattering rate of electrons. Here we assume that $\omega_p \gg \omega$ and $\omega \gg \nu$ or, at least, $\omega \gtrsim \nu$. These conditions are satisfied for many metals in IR and some part of the visible spectrum.

Considering a quasi-monochromatic incident laser pulse with the central frequency $\omega_0$ which satisfies the resonant condition (18), we calculate the amplitudes (16)–(17) at some shifted frequency $\omega = \omega_0 + \delta\omega$:

$$A(\omega) = C(\omega) = -\frac{\omega_0}{\delta\omega - i\nu\omega_0^2/2\omega_p^2} \frac{\omega_0^4}{4\omega_p^4} E_i(\omega)\delta\tilde{\varepsilon}. \tag{20}$$

Here we assumed that the spatial profile of the perturbation $\delta\tilde{\varepsilon}(x)$ is proportional to $E_z^\pm(x)E_t(x)$, so that $g = \alpha + \alpha_2 \cong 2\omega_p/c$. Below we show that this is valid in the case of quite fast interactions, before the diffusion significantly change the heat distribution inside the metal. In time domain, Eq. (20) can be rewritten as an equation for the envelopes:

$$\left(\frac{\partial}{\partial t} + \nu\frac{\omega_0^2}{2\omega_p^2}\right)\tilde{E}_z^\pm(t) = -\frac{\omega_0^6}{4\omega_p^5}\tilde{E}_i(t)\delta\tilde{\varepsilon}, \tag{21}$$

where $\tilde{E}_z^\pm(t)$ and $\tilde{E}_i^\pm(t)$ are slowly-varying time envelopes of SPP and of the incident pulse respectively. The growth rate of SPP is proportional to the incident field magnitude and to the current permittivity perturbation; the absorption of SPP takes time $\tau_a = \nu^{-1}\frac{2\omega_p^2}{\omega_0^2}$ which coincides with the well-known expressions for the SPP propagation length [16]. Thus, we obtained the equation of resonant excitation of counter-propagating SPP on given permittivity perturbation $\delta\varepsilon(x,z)$. Below we demonstrate the existence of positive feedback, so that the presence of SPP should increase the magnitude of $\delta\varepsilon$.

**Growth of the permittivity perturbation**. Total electric field inside the metal is the sum of electric fields of the transmitted wave and SPP. Electric field of SPP under the surface is almost tangential, so the total field $\mathbf{E}_\Sigma$ can be expressed as:

$$x < 0: \quad \mathbf{E}_\Sigma = Re\,(E_t e^{\alpha x} + 2E_z^\pm e^{\alpha_2 x}\cos k_\varepsilon z)e^{i\omega t}\mathbf{z}_0. \tag{22}$$

Let us consider the perturbation of permittivity caused by the periodically modulated electric field (22). Suppose that the medium is a metal with high free electron density $n_e$ which is expected to change relatively weak during the laser pulse action. At damaging fluences the electronic temperature reaches several eV which leads to several time increase of the collision frequency. For example, in Ref. [17], where the ultrafast heating of gold was studied, it was found that at incident laser fluences of about 0.5–1 J/cm$^2$, the density of free electrons increases only 30-50% while the collision frequency increases more than an order of magnitude.

In general case, comparable contributions to the increase of total electron collision frequency $\nu$ are made by electron-electron and electron-phonon scattering processes, which are mainly determined by the electronic and crystal lattice temperatures, respectively. For the sake of simplicity, here we use a linear approximation for the dependence of $\nu$ on the absorbed laser pulse energy per one electron $W_e$:

$$\nu(W_e) = \nu_0 + \xi\frac{W_e}{\hbar}, \tag{23}$$

where $\hbar$ is the Plank constant, $\nu_0$ is the initial collision frequency (for example, at room temperature), $\xi$ is a dimensionless constant, which expected to be of the order of 1 in the simplest theoretical model of Fermi liquid [18]. From recent experimental data on ultrafast heating of gold we can find that $\xi \approx 0.5$ [17].

When the incident laser pulse is acting on metal, equation of electron heating has the following form:

$$\frac{\partial W_e}{\partial t} = \nu \frac{e^2 |\mathbf{E}_\Sigma|^2}{2m\omega_0} \cong \nu \frac{e^2}{2m\omega_0} \left| E_t^2 e^{2\alpha x} + 4 E_t E_z^\pm e^{(\alpha+\alpha_2)x} \cos k_\varepsilon z \right|, \qquad (24)$$

where $e$ is the elementary charge; it is also taken into account that $E_t$ and $E_z^\pm$ are in phase in time domain according to Eq. (21) and that $E_z^\pm \ll E_t$. From Eq. (24) it follows that the presence of SPP field leads to the periodic spatial modulation of heating source $\sim e^{(\alpha+\alpha_2)x} \cos k_\varepsilon z$, which causes the same modulation of the collision frequency (23) and, finally, of metal permittivity (19). This proves the presence of positive feedback between the SPP excitation and $\delta\varepsilon$ growth.

From Eqns. (21) and (24), using definitions (19) and (23), we obtain the equation describing nonlinear growth of SPP magnitude:

$$\frac{\partial^2 \tilde{E}_z^\pm}{\partial t^2} + \nu \frac{\omega_0^2}{2\omega_p^2} \frac{\partial \tilde{E}_z^\pm}{\partial t} = \xi \nu \frac{\omega_0^4}{\omega_p^4} \frac{E_i^2 e^2}{\hbar m \omega_0^2} \tilde{E}_z^\pm. \qquad (25)$$

Initially, when the collision frequency perturbation is relatively small ($\nu_0 \gg \xi W_e / \hbar$), the SPP magnitude increases exponentially as $\tilde{E}_z^\pm = \tilde{E}_{z,0}^\pm e^{\Gamma_0 t}$, where the increment is given by

$$\Gamma_0 = \frac{\omega_0^2}{\omega_p^2} \sqrt{\xi \nu_0 \frac{E_i^2 e^2}{\hbar m \omega_0^2}}. \qquad (26)$$

For the parameters of gold ($\xi = 0.5$, $\varepsilon_0 \cong -26 - 1.85i$) and 100-fs laser pulse of 800 nm central wavelength and fluence of 1 J/cm$^2$, we found $\Gamma^{-1} \cong 50\ fs$. With the overall heating of electrons in the skin-layer, collision frequency $\nu$ increases more than 10 times [17], so the characteristic time of SPP growth decreases down to 10–15 fs. It means that during the action of an intense femtosecond laser pulse the magnitude of SPP will increase $10^1$–$10^3$ times due to the described instability.

Here we should note that the heat transport effects neglected in Eq. (24) will gradually change the permittivity profile $\delta\tilde{\varepsilon}(x)$ at the characteristic timescale of 200-300 fs (heat diffusion along the $z$-axis is several orders slower). This slightly decrease the instability increment. On the other hand, heat diffusivity also decreases with an increase of electronic temperature, which enhances the temperature difference between the neighboring near-surface regions. The influence of this effect on heat deposition and LIPSS formation was studied in Ref. [11].

Depending on the excitation conditions and on the material, SPP can be either fully absorbed during the laser pulse action or exists for several hundred femtoseconds after pumping. If the absorption time $\tau_a$ is less than the laser pulse duration, the distribution of deposited energy is described by Eq. (24), so the model predicts appearing of LIPSS with the period $2\pi/k_\varepsilon$. If $\tau_a$ is significantly longer, then SPP will be absorbed mostly after the laser pulse reflection, so the absorption profile will follow the SPP standing wave intensity $\propto \cos^2 k_\varepsilon z$ and the LIPSS period is expected to be of about $\pi/k_\varepsilon$.

As it follows from Eq. (21), $\tau_a$ decreases with the increase of collision frequency during the electron heating ($\tau_a$ changes approximately from 370 fs to 30 fs for the parameters of gold listed above). This will limit the exponential growth. According to Eqns. (21) and (19), saturated magnitude of SPP does not depend on $\nu$ when $\nu \gg \nu_0$ and can be expressed as

$$\tilde{E}_{z,sat}^{\pm} = i\frac{\omega_0}{2\omega_p}\tilde{E}_i \cong E_t/4. \qquad (27)$$

Thus, in the saturation regime the internal electric field given by Eq. (22) varies 3 times along $z$-axis (from $0.5E_t$ to $1.5E_t$), so the electric field intensity $|\mathbf{E}_\Sigma|^2$ changes almost an order of magnitude. Strictly speaking, this takes us beyond the limits of full applicability of the perturbation theory, so that Eq. (27) should be treated as an estimation proving that the magnitude of SPP electric field can be comparable to the magnitude of the transmitted electric field inside the medium $E_t$. Nevertheless, the regime of saturation seems to be the most probable in the experiments on single-pulse LIPSS formation where the electrons in the skin-layer are heated up to several eV and the collision frequency becomes comparable to $\omega_0$.

SPP diffraction on the permittivity grating $\delta\varepsilon(z)$ also shortens their lifetime. Due to this effect SPPs are transformed to the reflected optical wave propagating normally to the surface. Corresponding lifetime can be found by solving the diffraction problem similar to the problem of direct transformation (Eqns. (5)–(21)); according to our calculation, diffraction lifetime can be estimated as $t_{diff} \cong 16(-\varepsilon_0)^{7/2}/\omega\delta\varepsilon^2$, which gives ~3–4 ps for the parameters discussed above. So, under the chosen conditions the effect is negligible.

To conclude, we showed that normal reflection of an intense laser pulse from a metal surface is an unstable process. Excitation of seed surface plasmon-polaritons at any random permittivity perturbations or surface irregularities leads to the formation of a periodic interference pattern and, consequently, to the periodic heating of the medium and permittivity modulation. In turn, this periodic modulation of permittivity occurs to be resonant for further enhancement of conversion of the incident wave into the pair of counter-propagating SPPs. This positive feedback leads to several order growth of SPP magnitude during the action of the femtosecond laser pulse.

Described instability seems to be promising for the explanation of LIPSS formation in the single-pulse or few-pulse regime (when the theoretical models based on the inter-pulse feedback cannot interpret the appearance of LIPSS satisfactory). For the parameters of gold, the instability becomes faster than the typical laser pulse duration at the fluences of about ~0.5–1 J/cm$^2$ which is the experimentally observed threshold of single-pulse LIPSS formation [12]. As it follows from the developed theory, spontaneous growth of SPP due to the thermal nonlinearity saturates when the magnitude of the SPP standing wave reaches ~1/2 of the transmitted electric field inside the metal. In contrast to previous studies, fast exponential growth and saturation of SPP make the presented model almost insensitive to the amplitude of initial surface roughness. Notice that the laser-induced formation of periodic ripples inside the glass was recently interpreted in Ref. [19] as a consequence of the instability of a flat wave reflection from an ionized dielectric layer.

In the case of few-picosecond laser pulses (or longer) the discussed mechanism of decay into SPPs is still relevant, however the crystal lattice heating and heat diffusion should be taken into account explicitly. The total electron collision frequency is also expected to increase periodically, but the dominant contribution should come from electron-phonon scattering rate which grows linearly with the lattice temperature $\nu_{e-ph} \propto T_l$.

**Acknowledgement.** The work was supported by the Ministry of Science and Higher Education of the Russian Federation (project #075-15-2020-790) and by the Theoretical Physics and Mathematics Advancement Foundation "BASIS" (project #19-1-4-61-1). The author is also grateful to V. A. Mironov and V. B. Gildenburg for fruitful discussions.